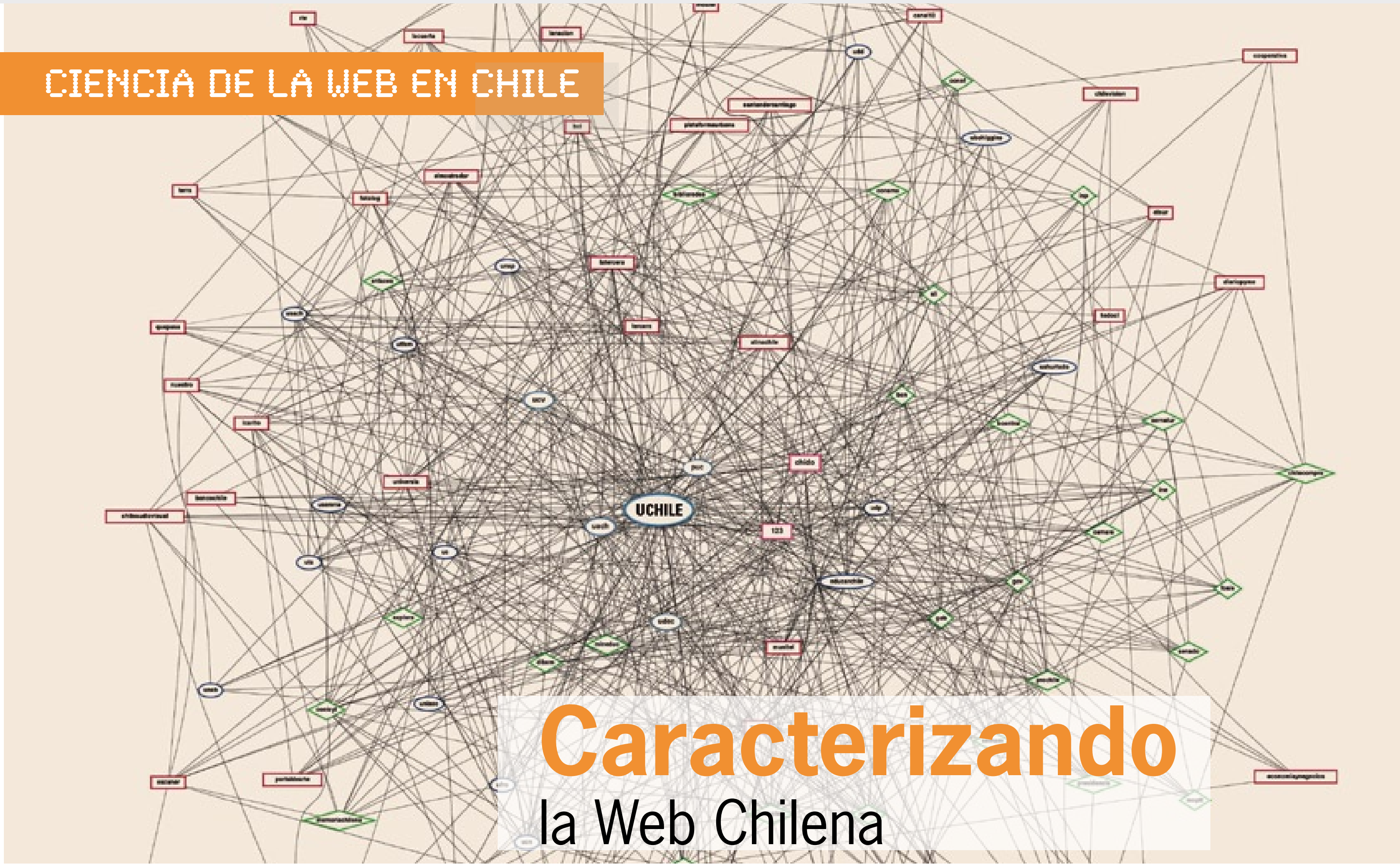

CIENCIA DE LA WEB EN CHILE

# Caracterizando
## la Web Chilena

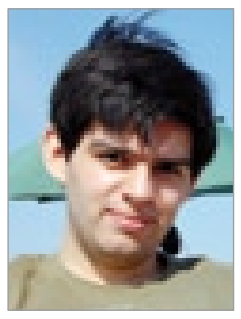

**Eduardo Graells**
*Estudiante de Magíster en Ciencias mención Computación, DCC, Universidad de Chile. Ingeniero Civil en Computación de la misma Universidad.*
*egraells@dcc.uchile.cl*

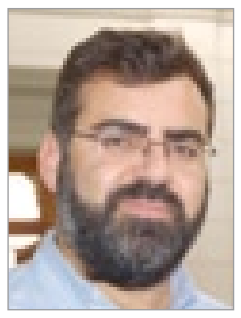

**Ricardo Baeza-Yates**
*Profesor Titular, DCC, Universidad de Chile. Profesor ICREA Asociado de la Universitat Pompeu Fabra, Barcelona. Ph.D. en Computer Science, University of Waterloo. Vicepresidente de Yahoo! Research para Europa, Medio Oriente y Latinoamérica.*
*rbaeza@dcc.uchile.cl*

## INTRODUCCIÓN

La Web es más que un simple conjunto de documentos en distintos servidores, ya que existen relaciones de información entre los documentos mediante los enlaces que se establecen entre ellos. Esto mejora la experiencia de navegación de los usuarios porque los ayuda a encontrar información, y ayuda a los programas que recorren la Web que buscan enlaces de nuevos documentos dentro del contenido de los documentos ya revisados. De ese modo funcionan los motores de búsqueda: permiten encontrar documentos que contengan ciertas palabras claves, y dichos documentos se encontraron siguiendo enlaces. Los resultados de la búsqueda se presentan ordenados de acuerdo a la cantidad de enlaces que reciben (entre otros parámetros que afectan el ordenamiento). Esto permite interpretar el número de enlaces que recibe un documento como una medida de su calidad, porque en general una página enlaza a otras similares.

La Web Global se puede considerar como un gran grafo que tiene una estructura que se puede clasificar como *red libre de escala*, que, al contrario de las *redes aleatorias*, se caracteriza por una distribución dispareja de enlaces, en la que los nodos altamente enlazados actúan como centros que conectan muchos de los otros nodos a la red. Analíticamente, este comportamiento disparejo se puede expresar mediante una ley de potencias (*powerlaw*):

$$\textit{frecuencia} \approx kx^{-\theta}$$

donde $k$ es una constante que depende del contexto, $x$ es el número de enlaces y $-\theta$ es el parámetro de la distribución. Esto quiere decir que la distribución de los enlaces es muy sesgada: unas pocas páginas reciben muchos enlaces mientras que la mayoría recibe muy pocos o incluso ninguno. En este artículo se muestra que dicha distribución se puede aplicar a una gran cantidad de aspectos de la Web Chilena, validando la consideración de la Web Global como similar a una red libre de escala, porque estas son auto-similares: una pequeña muestra mantiene características de la red completa.

La Web Chilena se define como el conjunto de sitios cuyo dominio de primer nivel es .cl, o que estén hospedadas en un servidor cuya dirección IP está asociada a Chile. Entre los años 2000 y 2007 sus características han sido estudiadas por el Centro de Investigación de la Web (CIW) y el buscador TodoCL[1]. Esta caracterización se realiza mediante una recolección o *snapshot* de la Web Chilena en un momento particular en el tiempo. La ultima recolección se realizó en septiembre del año 2007, y luego en octubre del mismo año, se realizó una recolección considerando solamente los sitios que tenían enlaces entrantes o salientes, con el fin de obtener una mejor caracterización de los dominios y sitios con más enlaces.

A pesar de estar estudiando un subconjunto acotado de la Web Global, las propiedades que se han encontrado en la Web Chilena son similares a las globales en términos de las distribuciones mencionadas.

A continuación describimos la colección y luego presentamos los resultados a nivel de páginas, sitios y dominios, terminando con las conclusiones.

> La cantidad de servidores que no entregan información no es despreciable, aunque en general se puede concluir que en los servidores las tecnologías de código abierto superan ampliamente a las tecnologías propietarias.

## COLECCIÓN DE LA WEB CHILENA 2007

Habiendo definido qué es la Web Chilena, es necesario recorrerla para obtener los documentos y sitios que la componen. Para realizar la colecta se utilizó el *crawler* WIRE 0.14[2], que a partir de una lista inicial de sitios (semillas o *seeds*) comienza a descargar sus documentos, con el fin de almacenarlos, analizar su contenido y encontrar enlaces a más sitios que son agregados a la lista de sitios por descargar. El proceso se repite hasta que se han descargado todos los documentos públicos posibles, cuando se han agotado algunos parámetros (como el espacio en disco o el límite de documentos a descargar), o cuando se ha llegado a un punto en el cual sólo se están descargando documentos redundantes (situación que ocurre debido a la generación de páginas y URLs dinámicas).

Para la colección que se utilizó en la caracterización del año 2007, se utilizó un computador con una CPU Intel Pentium IV de 3 GHz, 1 GB de memoria RAM y sistema operativo Ubuntu 7.04. El Cuadro 1 resume las características principales de la colección.

| Páginas Web | 9.637.801 |
|---|---|
| Texto en total | 135,76 [GB] |
| Texto promedio por página | 14,77 [KB] |
| Sitios Conocidos | 200.000 |
| Sitios Recolectados | 111.374 |
| Páginas promedio por sitio | 86,53 |
| Texto promedio por sitio | 1,24 [MB] |
| Dominios Conocidos | 190.577 |
| Dominios Recolectados | 104.409 |
| Sitios promedio por dominio | 1,07 |
| Páginas promedio por dominio | 92,31 |
| Texto promedio por dominio | 1,33 [MB] |

**Cuadro 1** Resumen de estadísticas de la colecta.

## CARACTERÍSTICAS DE

## LOS DOCUMENTOS

La caracterización de la Web Chilena contempla 9.637.801 documentos o páginas Web que fueron descargadas desde la Web pública y/o visible. Una página promedio pesa 14,77 KB sin considerar imágenes u otros contenidos multimedia incrustados en ella, aunque los enlaces a archivos multimedia o documentos de texto son registrados para su posterior análisis. En el tamaño de las páginas es donde se encuentra la primera ley de potencias: la distribución de tamaño de los documentos versus la fracción de los documentos sigue una ley de potencias con parámetro -3,56 para páginas de más de 40 KB, y de -0,82 para páginas entre 11 y 40 KB.

Al descargar cada página, el servidor Web da a conocer la fecha en la cual dicha página se modificó por última vez. Este dato nos permite modelar la distribución de la edad de los documentos, en la cual nos damos cuenta que en los 12 meses previos a la recolección se actualizó un 19% de los documentos de la colección. Naturalmente también hay documentos que no se actualizan hace mucho tiempo, pero esa cantidad está acotada por los documentos encontrados en las colectas anteriores. Esta descripción es propia de una ley de potencias: la distribución de la edad de los documentos se puede aproximar con una ley de parámetro -1,27.

[1] http://www.todocl.cl

[2] http://www.cwr.cl/projects/WIRE

De las páginas descargadas, 34% de ellas son dinámicas; páginas generadas en el momento de ser solicitadas sin que existieran previamente. Esto es muy común en la actualidad; el contenido de un sitio se almacena en una base de datos y las páginas en realidad son programas que leen dicho contenido al momento de ser solicitada la página por el usuario. Dichos programas consideran distintas variables, como aquellas entregadas en la URL o una posible autenticación previa del usuario en el sitio. La aplicación más usada para generar estas páginas es PHP[3], una tecnología de código abierto, que tiene un 79,36% de participación. Le sigue la tecnología ASP, propietaria y de plataforma restringida, con un 18,07%.

Respecto a enlaces a documentos no HTML, estos se pueden agrupar por tipo:

- **Texto:** 1,5 millones de enlaces. Los formatos más populares son PDF (56,74%), XML (26,69%) y DOC (6,51%).
- **Imagen:** 100 millones de enlaces. Los formatos más populares son GIF (77,26%), JPG (18,26%) y PNG (4,45%)
- **Audio:** 166 mil enlaces. Los formatos más populares son WMA (40,29%) y MP3 (39,23%).
- **Vídeo:** 35 mil enlaces. Los formatos más populares son WMV (49,59%), QT (18,20%), MPEG (10,65%) y RM (10,54%).

## CARACTERÍSTICAS DE LOS SITIOS

La recolección se inició con una lista de seeds o semillas que contenía todos los dominios .cl registrados, así como sitios no .cl que se conocían de colectas anteriores. La lista de dominios la provee NIC Chile gracias a un acuerdo de investigación. A cada uno de esos dominios se les agrega el prefijo www, ya que en general es correcto asumir que las direcciones http://sitio.cl y http://www.sitio.cl apuntan al mismo sitio.

En el proceso de recolección se llegó a conocer un total de 200.000 direcciones de sitios, aunque solamente 111.374 pudieron ser recolectados. Aquellos que no pudieron ser recolectados no tenían una dirección IP asociada al momento de realizar la recolección. Así, un sitio tiene en promedio 86,53 páginas y un contenido HTML total de 1,24 MB. Sin embargo, estas distribuciones también presentan leyes de potencia, lo que indica que existen muchos sitios con pocas páginas o contenido y pocos sitios que agrupan una gran cantidad de páginas y del tamaño total de la Web Chilena: sólo un 7% de los sitios tiene el 90% de los documentos, y un 14% de los sitios contiene el 99% del total del contenido. Los parámetros de dichas leyes de potencia son -1,84 para el número de páginas por sitio y -1,64 para el tamaño de los sitios.

Los enlaces de entrada y salida que tiene un sitio también son estudiados. Se definen como grado interno y grado externo: el grado interno de un sitio *S* es el número de sitios distintos que enlaza a *S*, y el grado externo de *S* es el número de sitios distintos enlazados por *S*. Es decir, dentro del grado interno es indiferente si distintas páginas dentro del mismo sitio son enlazadas por otro o enlazan a otro, en ambos casos, todos esos enlaces sólo incrementan en uno el grado correspondiente. Este esquema de enlaces forma un grafo de sitios de nido como *Hostgraph*. En el Hostgraph la distribución del grado interno y externo para los sitios es muy sesgada, ya que hay pocos sitios que reciben una gran cantidad de enlaces y muchos que reciben pocos o incluso ninguno. Las leyes de potencias que modelan estas distribuciones tienen parámetros -2,16 (grado interno) y -2,32 (grado externo)[4].

El Cuadro 2 muestra los 5 sitios más destacados en el número de páginas tamaño en GBs, grado interno y grado externo. Los sitios con mayor grado interno han mantenido su posición constantemente a lo largo de los años, mientras que los sitios con más páginas y contenido cambian cada año.

| PÁGINAS | CONTENIDO (GB) | GRADO INTERNO | GRADO EXTERNO |
|---|---|---|---|
| www.autovia.cl (22.825) | www.suena.cl (1,67) | www.sii.cl (542) | www.chido.cl (1.253) |
| www.b2.cl (22.473) | www.amazon.cl (1,55) | www.uchile.cl (398) | www.fotolog.cl (523) |
| www.ais.cl (22.100) | www.planetashile.cl (1,15) | www.mineduc.cl (374) | www.atinachile.cl (416) |
| www.kontent.cl (21.613) | listados.deremate.cl (0,91) | www.meteochile.cl (335) | www.todocl.cl (352) |
| www.madness.cl (21.244) | www.b2.cl (0,85) | www.corfo.cl (290) | www.webs.cl (292) |

**Cuadro 2** Sitios destacados en las distintas variables analizadas.

## ESTRUCTURA DE LA WEB CHILENA

En un grafo, una componente fuertemente conectada (SCC por *Strongly Connected Component*) es aquella en la que se puede llegar desde un nodo hasta cualquier otro siguiendo las aristas en el grafo, respetando la dirección de éstas. Este análisis se puede aplicar al hostgraph, y es así como se encuentra una SCC gigante y una cantidad menor de componentes fuertemente conectadas más pequeñas (incluyendo SCCs de tamaño 1). El tener una SCC gigante es un signo típico de redes libres de escala.

La SCC gigante de la Web Chilena tiene 6275 sitios, y puede considerarse como el punto de partida desde el cual se define la estructura de la Web Chilena.

[3] http://www.php.net

[4] Estos valores corresponden a la colecta de septiembre. Para la realizada en octubre, los valores son -1,83 y -1,84 para los grados interno y externo, respectivamente.

En base a la conectividad que tienen los sitios con aquellos presentes en la SCC gigante, se pueden definir las siguientes componentes:

- **MAIN**, los sitios en la componente fuertemente conexa.
- **OUT**, los sitios que son alcanzables desde MAIN, pero que no tienen enlaces hacia MAIN.
- **IN**, los sitios que pueden alcanzar a MAIN, pero que no tienen enlaces desde MAIN.
- **ISLAS**, sitios desconectados de los demás en términos de enlaces.
- **TENTÁCULOS**, sitios que sólo se conectan con IN o OUT, pero en el sentido inverso de los enlaces.
- **TÚNEL**, una componente que une las componentes IN y OUT sin pasar por MAIN.

La componente MAIN se extiende en las siguientes subcomponentes:

- **MAIN-MAIN** son los sitios que pueden ser alcanzados directamente desde la componente IN o que pueden alcanzar directamente la componente OUT.
- **MAIN-IN** son los sitios que pueden ser alcanzados directamente desde IN pero no están en MAIN-MAIN.
- **MAIN-OUT** son los sitios que pueden alcanzar directamente a OUT pero no pertenecen a MAIN-MAIN.
- **MAIN-NORM** son los sitios que no pertenecen a las subcomponentes definidas anteriormente.

La Figura 1 muestra una representación gráfica de la estructura descrita y la distribución de sitios y páginas a través de las componentes. La componente ISLAS contiene la mayor cantidad de sitios, pero la que tiene más páginas es MAIN (en particular MAIN-MAIN). Esta estructura permite encontrar relaciones entre los sitios que pertenecen a cada componente: usualmente en IN se encuentran portales o sitios de inicio en la red, que buscan enlazar a sitios importantes; en OUT se encuentran sitios que buscan recibir enlaces pero que no entregan enlaces a otros sitios; en MAIN suelen haber sitios interconectados entre sí como pueden ser sitios de universidades y del gobierno. En la componente ISLAS se encuentra la mayoría de los sitios con una única página recolectada, aunque es posible que si se consideran las páginas que no se recolectaron en esos sitios se encuentren enlaces que permitan llevar a algunos de los sitios a IN o a TENTÁCULOS. Adicionalmente, en MAIN se encuentra la mayoría de los enlaces a documentos de texto en formato no HTML.

## CARACTERÍSTICAS DE LOS DOMINIOS

Si bien se conocen 190.577 dominios distintos, solamente se pudieron recolectar 104.409, aunque de éstos se pudieron contactar 117.700. Esto quiere decir que hay dominios que tienen una dirección IP asociada pero no tienen ninguna página (o bien tienen páginas, pero éstas son privadas), ya que al tratar de conectarlos entregan códigos de error (como pueden ser *404 - Not Found o 403 - Forbidden*). En total se encontraron 14.477 direcciones IP. La distribución de los dominios en estas direcciones también sigue una ley de potencias, de parámetros -0; 35 en su parte inicial y -1; 37 en su parte central. La distribución es tan sesgada que 2 direcciones tienen más de 1.000 dominios, y 9.026 direcciones tienen solamente 1 dominio cada una.

Para cada dirección IP se le pidió al servidor información sobre el software instalado mediante un *Request HTTP*. Como resultado se obtuvo lo siguiente:

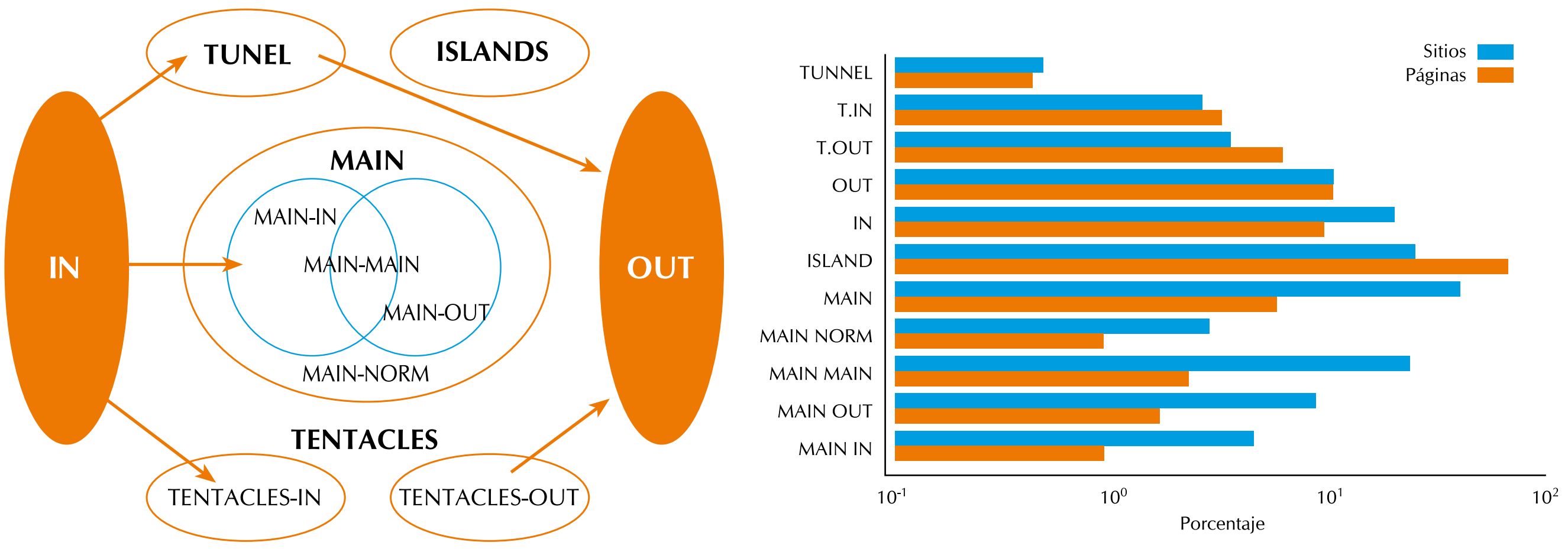

**Fig. 1** A la izquierda, una representación gráfica de la estructura de la Web Chilena. A la derecha, la distribución de sitios y páginas en las componentes (en escala logarítmica).

- **Sistema Operativo:** 43,21% no entrega información, 38,67% usa GNU/Linux y Unix, 18,12% usa Microsoft Windows.
- **Servidor Web:** 38% no entrega información, 43% usa Apache, 18,12% usa Microsoft IIS.

La cantidad de servidores que no entregan información no es despreciable, aunque en general se puede concluir que en los servidores las tecnologías de código abierto superan ampliamente a las tecnologías propietarias. Esto es coherente con los resultados obtenidos en la identificación de tecnologías para páginas dinámicas.

Respecto a las características analizadas en las secciones anteriores, el Cuadro 3 muestra los 5 dominios más destacados en cantidad de sitios, contenido y grado interno. En general, el comportamiento de los dominios es más estable que el de los sitios, aunque también se pueden encontrar anomalías, como puede verse en los dominios más enlazados, donde existen tres dominios que apuntan al mismo sitio, correspondiente a una protección de dominios.

En base a los enlaces entre dominios se ha creado una representación gráfica de la Web Chilena, visible en la Figura 2. Para esta representación se eligieron los 31 dominios más conectados entre sí, considerando de la lista de dominios más enlazados solamente aquellos que tenían sitios en la componente MAIN MAIN. Los dominios son representados como nodos enlazados por una línea cuyo grosor y color muestra la cantidad de enlaces entre ellos (mientras más oscuro y grueso, hay una mayor cantidad de enlaces). Se dividen en tres grupos: comerciales (rectángulos), de instituciones educacionales (elipses) y de gobierno (rombos). En la imagen se aprecia que dominios del mismo tipo tienden a estar más cercanos entre sí.[5]

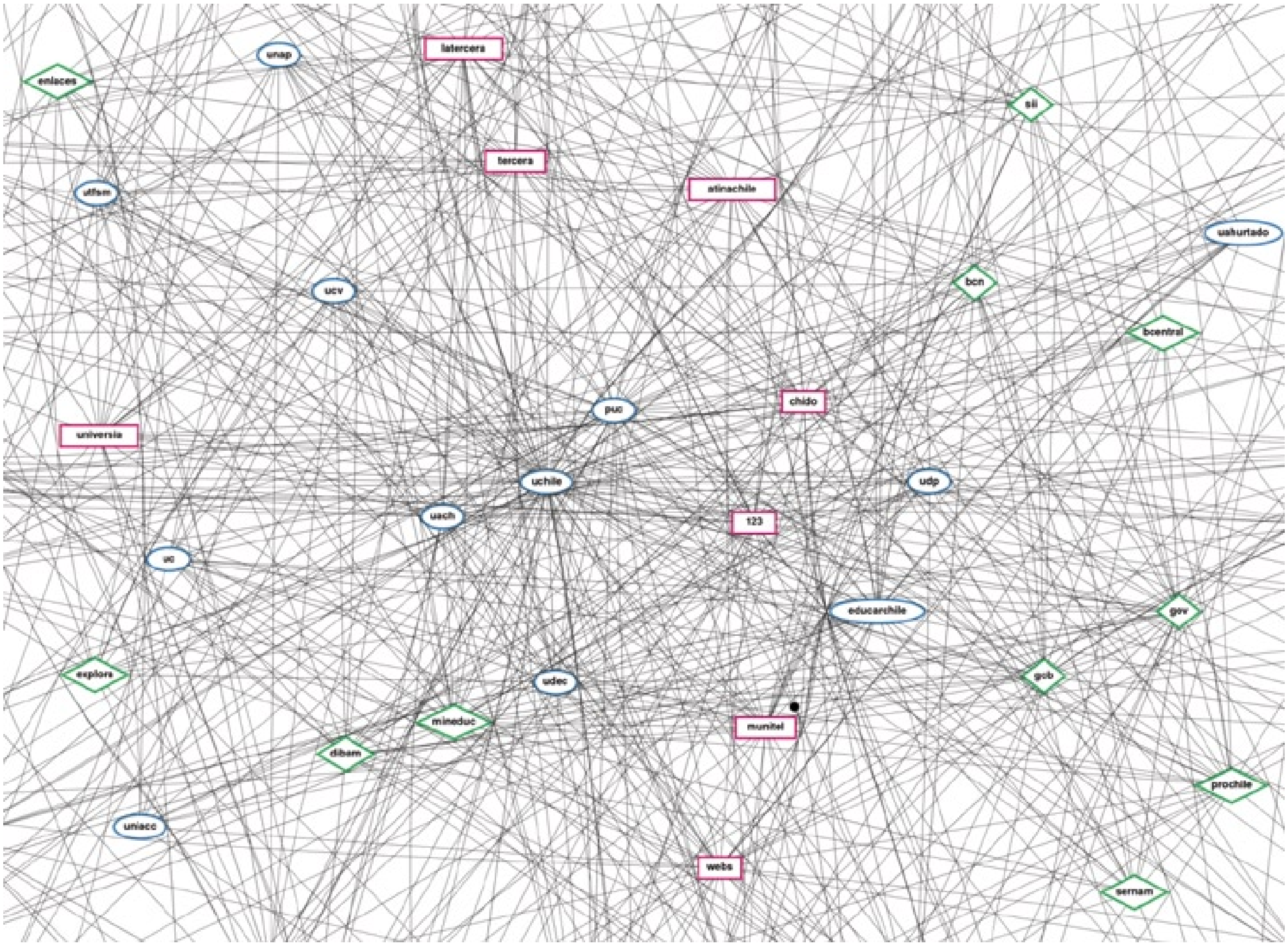

**Fig. 2** Una visión gráfica de los 31 dominios más conectados entre sí de la Web Chilena.

[5] En el reporte técnico "Características de la Web Chilena 2007" se observa esta imagen cabalmente, con más de 100 dominios.

# CONCLUSIONES

La Web Chilena ha cambiado bastante respecto a los últimos años y, a pesar de estar en constante cambio, sigue manteniendo una estructura similar a la encontrada en años anteriores. El crecimiento en la cantidad de documentos recolectados desde el año 2006 es notorio, desde 7; 4 millones a 9; 6 millones, lo cual es consecuente con la cantidad de documentos creada o actualizada en los últimos 12 meses. La distribución de los documentos en diferentes análisis se puede ajustar a leyes de potencias, verificando el modelo de redes libres de escala.

Aunque se conocía la dirección de 200;000 sitios, sólo se pudieron recolectar cerca de 111;000. El análisis de algunas características de los sitios también presenta leyes de potencias: la distribución de documentos por sitios, la del contenido y la de enlaces entre sitios. Además, los sitios que reciben más enlaces se han mantenido a lo largo de los años, y destacan por ser sitios del gobierno, de instituciones educacionales o de medios de comunicación. La macroestructura de la web también presenta características importantes: aunque solamente un 5% de los sitios recolectados está fuertemente conectado entre sí, estos sitios tienen el 39% del total de las páginas. A su vez, un 65; 26% de los sitios está aislado de los demás, y contienen cerca del 24% del total de las páginas.

| SITIOS | CONTENIDO (GB) | GRADO INTERNO |
|---|---|---|
| portalciudadano (690) | turismo-viajes (3,04) | uchile (1.300) |
| uchile (374) | suena (1,68) | nameaction, backorder, snapnames (906, 904, 902) |
| scd (352) | deremate (1,63) | gov (653) |
| loquegustes (342) | amazon (1,55) | puc (550) |
| boonic (267) | mercadolibre (1,55) | sii (542) |

**Cuadro 3** Dominios destacados en las distintas variables analizadas. Los dominios nameaction, backorder y snapnames son *mirrors*.

La distribución de direcciones IP para los dominios también se ajusta a una ley de potencias. En estas direcciones se estudió la tecnología que utilizaba el servidor y, en las que entregaron información, se encontró que tanto en el sistema operativo como en el servidor utilizado las tecnologías de código abierto tienen mayor presencia.

Se concluye que la caracterización de una Web Nacional, en este caso la Web Chilena, permiten, además de establecer un modelamiento de la Web en términos matemáticos o analíticos, tener datos concretos que sirven de base para estudios de usabilidad, de mercado y de minería de datos, entre otros. Lo que se ha realizado es una captura de un instante particular de la existencia de la Web, cuya representatividad no se puede poner en duda al ver la constancia que se ha tenido durante los pasados años y los resultados similares vistos en estudios aplicados a otras Web Nacionales.BITS

# REFERENCIAS